\documentclass{mem}
\usepackage{natbib}
\usepackage{txfonts}
\usepackage{balance}
\usepackage{graphicx}
\usepackage[a4paper]{hyperref}
\idline{75}{282}
\begin{document}
\def\teff{$T\rm_{eff }$}
\def\kms{$\mathrm {km s}^{-1}$}

\title{
Galaxies and the Local Universe with WFXT
}

   \subtitle{}

\author{
G. \,Trinchieri\inst{1} 
\and A. \, Wolter\inst{1}
          }

  \offprints{G. \,Trinchieri }

\institute{
INAF --
Osservatorio Astronomico di Brera, Via Brera 28,
I-20121 Milano, Italy\\
\email{ginevra.trinchieri@brera.inaf.it}
}

\authorrunning{Trinchieri \& Wolter }

\titlerunning{Galaxies and the local Universe}

\abstract{
Galaxies are essential building blocks in the Universe. However they
are faint and complex X-ray sources and require high performance instrumentation to be properly studied. Yet they are fundamental for our understanding of the Universe, and a detailed knowledge of the local structures is mandatory to explain
the deep and far Universe. We make a few examples, and discuss how
well suited WFXT is to address this issue.
\keywords{X-ray: galaxies -- galaxies: spiral --  galaxies: general -- galaxies: ISM
}
}

\maketitle{ }

\section{Why devote time and efforts to local normal galaxies?}

Galaxies are relatively faint and complex
X-ray sources: their luminosities range between
L$_{X} \sim 10^{39}$ erg/s  and L$_{X} \sim 10^{41}$ erg $s^{-1}$, and their emission is due to several distinct components, whose relative importance is related to galaxy parameters such as morphology and evolutionary stage. Due to their complexity, our understanding of their X-ray properties is not as advanced as for other astronomical objects, and is limited to systems in the local universe. 
Nonetheless, they are fundamental if we aim at understanding the Universe at large, for no other reason that they have been recognized as  the dominant component in the extragalactic
sky at very low fluxes \cite[see][for a recent review]{Georg06}.
Therefore, only a precise determination of the
key components in local galaxies allows us to predict their behaviour at
high redshifts and/or to study their evolution.

The advantages in a detailed  study of local galaxies are manifold: \\
a) The ``same'' distance to all sources in a galaxy allows an investigation 
of specific source classes  more homogeneously. 
b) A few source classes can {\bf only} be studied outside Milky Way (e.g. ULXs, SSS, hot halos). 
c) It returns a better understanding of population of the Milky Way
where sources can be studied in more detail. 
d) It represents a baseline for comparison with the more distant universe: today's
galaxies are the end product of evolution and as such are
important traces of the distant Universe that we can investigate 
in detail. 
Only a precise knowledge of their properties will allow us to  properly interpret the ``unresolved'' emission from distant objects. 

\noindent In spite of the efforts to understand local Universe objects, a real break-through requires instruments with the following characteristics: 
$\bullet$
Large field of view: local objects have large angular size.  Given their complexity, it is important also to be able to determine the background locally, to avoid normalization and spectrum issues.  \\
$\bullet$ Very good spatial resolution.  This is required to separate different components, to minimize  reciprocal contamination, and to study individual sources in detail.  \\
$\bullet$ Sensitivity.  Given their faint nature, galaxies are  at low fluxes: 
at the limit of the Medium Survey, f$_{x} \sim 10^{-15}$ erg cm$^{-2}$ s$^{-1}$, they will all be detected out 
to D=90 Mpc.  Only the brightest ones ($ \sim 10^{41}$ erg s$^{-1}$) out to z$\sim$0.2. \\
$\bullet$ Broad energy band.  Different components have significantly different spectral characteristics, and their contributions dominate different regions of the spectrum.  In order to be able to separate them spectrally, a broad band is essential. \\
$\bullet$ ``Reasonable'' spectral resolution.  While a superb spectral resolution would be desirable, all other constraints would make this too demanding.   Nonetheless,  spectral analysis is needed to characterize different components and provides an excellent alternative in separating contributions from different sources when the spatial resolution is not adequate. \\

\noindent Most of the  above requirement are indeed met by WFXT, at least to a reasonable compromise to be useful for detailed studies of the local Universe.  The spatial resolution (required or goal) would not allow us to address some of the outstanding issues outlined below (Chandra resolution or better would be required), but a resolving power of 1-2$''$ would be enough to tackle many of them. 
 
\section{Outstanding issues where WFXT can contribute}

\begin{figure*}[ht!]
\resizebox{\hsize}{!}
{\includegraphics[width=8.5cm]{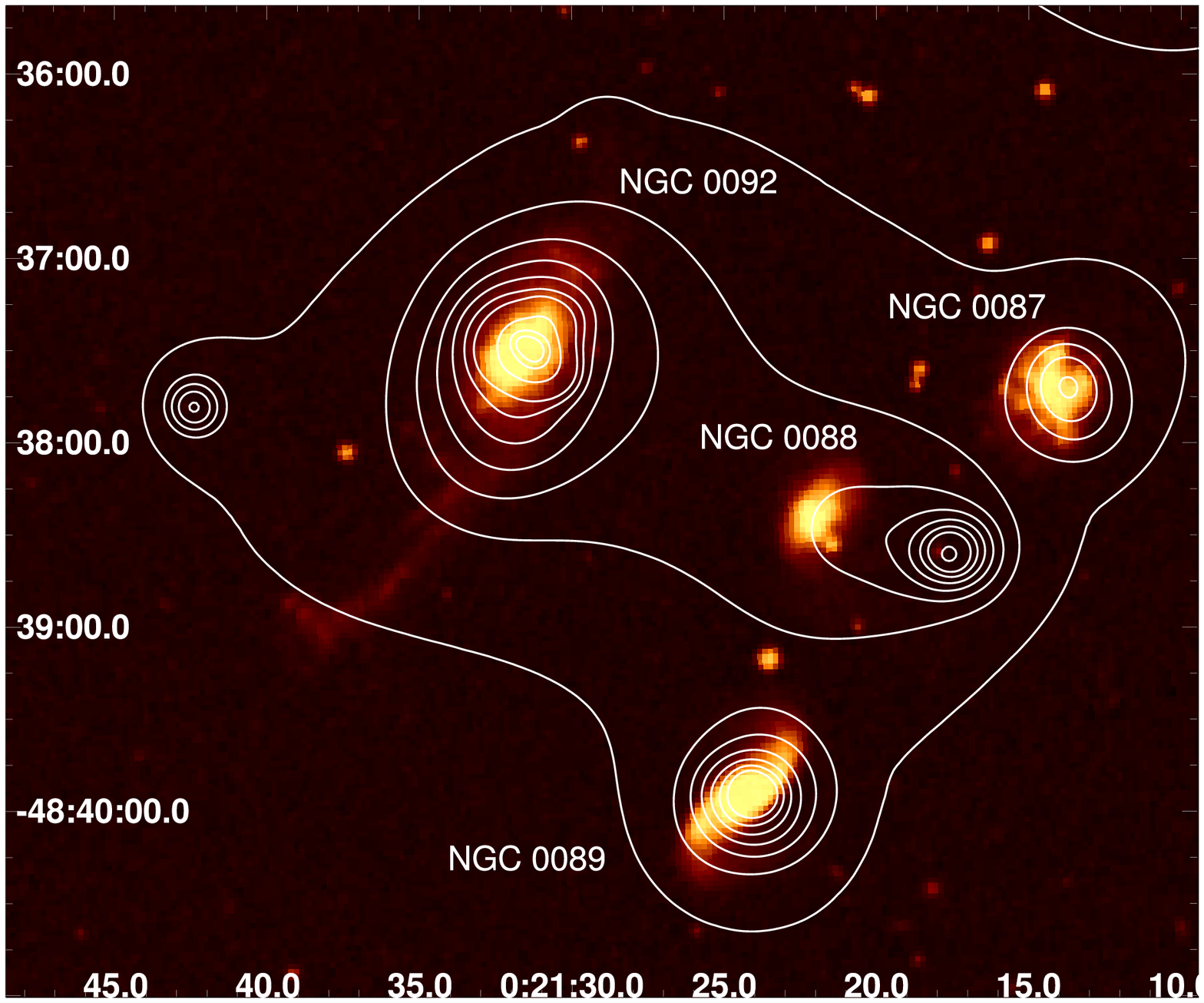} 
\includegraphics[clip=true,angle=90,width=13cm]{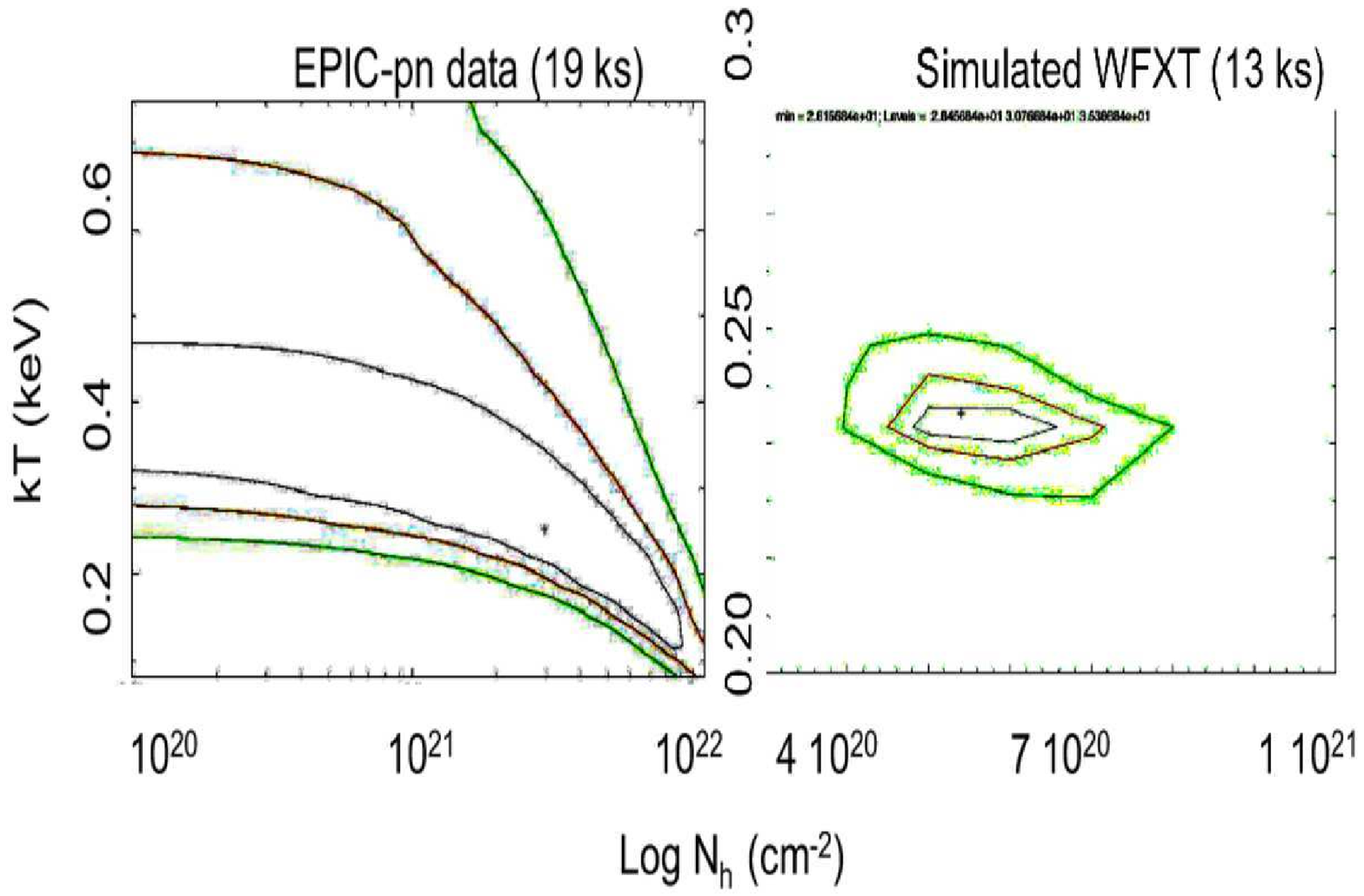}

}
\caption{\footnotesize
EPIC-pn observations of SCG0018-4854 \citep{Tri08} have identified a hot ISM in this spiral-dominated system ({\it left panel}).  However, its spectral properties are poorly determined. A WFXT observation of similar duration would significantly reduce the parameter space  ({\it  middle and right panels}). 
}
\label{SCG}
\end{figure*}

Several questions related to our understanding of the X-ray properties of galaxies still need to be addressed and properly studied. 
We list some here, and discuss a few in detail below, in context with WFXT characteristics. \\
{\bf
Luminosity Function of normal galaxies.}  This has  never properly been done on large enough samples to be able to distinguish them in different classes that reflect intrinsic characteristics.  We discuss this better in \S~\ref{XLF} \\
{\bf
What is the nature of ULXs?} This extreme class of sources can only be studied in external galaxies.  Current samples are limited, but WFXT will provide a reasonably large sample of sources in  diverse galaxies (see \S~\ref{ULX}). \\
{\bf
Dwarf galaxies}  This class has been so far neglected, largely due to the difficulty related to their observation.  A survey mode at reasonable depth is the ideal tool to  collect enough examples, and the large FoV will allow us to examine those around bigger galaxies. \\
{\bf
Hot gas in early type galaxies:} how much and why only in some of them? 
A large spread in the $\rm L_{X-gas} - L_B$ plane is observed,  and attempts at interpreting it in light of intrinsic galaxy properties have been so far only partially successful. Emission of galaxies located at group centers is hard to separate from that of the environment and requires special attention in the interpretation in the context of galaxy properties. Even a careful selection of galaxies as a function of environment has not led to conclusive results \citep{Memola09, Mulchaey10}.   Investigating this with larger and well studied samples might help to better assess what the relevant parameters are (see Pellegrini, this conference).\\
{\bf
Hot gas in the local universe}: how much? and what physical state? 
We have seen the presence of instabilities such as hot/cold bubbles and cavities in clusters and now also around bright or central group galaxies.
Large halos have been detected around spiral galaxies, and in the intergalactic space.  Studying these phenomena on a 
large number of galaxies will allow us to put the presence and frequencies
of these effects on a statistical ground and to correlate them with
intrinsic properties (see \S~\ref{gas})\\
\noindent{\bf
Young spiral-dominated groups.}  As part of our understanding of the presence and amount of hot gas in the local universe, a new and largely unexplored field is that of poor groups, in particular those at a young dynamical age. As a whole, groups could be regarded as the fainter end  of the cluster class,  and their properties 
interpreted in this context. However, so far very little is known
about the X-ray properties of spiral-only compact groups, which could
represent the beginning of an evolutionary sequence that would lead
them into more relaxed systems, dominated by an early-type
galaxy population, through a multiplicity of mechanisms such as
mergers of member galaxies, accretion of new gas from external
reservoirs, infall of new galaxies and heating of the intra-group
medium through dynamical friction and AGN feedback. A hot ISM in these
systems is very hard to observe because it appears to be significantly
fainter and possibly cooler than in more evolved systems
(\citealt{belsole, Tri08}).  However determining its presence could
have a significant impact on our understanding of these early stages
of group evolution.  Moreover we expect to encounter more at higher z,
therefore making their interpretation more challenging if we lack a
good knowledge of their properties locally. Fig.~\ref{SCG}
demonstrates the improvement we can expect on determining the presence
and the characteristics of a hot ISM in these systems.

\section{WFXT and the local Universe: how good a match is it?}

\begin{figure*}[ht!]
\resizebox{\hsize}{!}
{\includegraphics[clip=true,width=13cm,height=6cm]{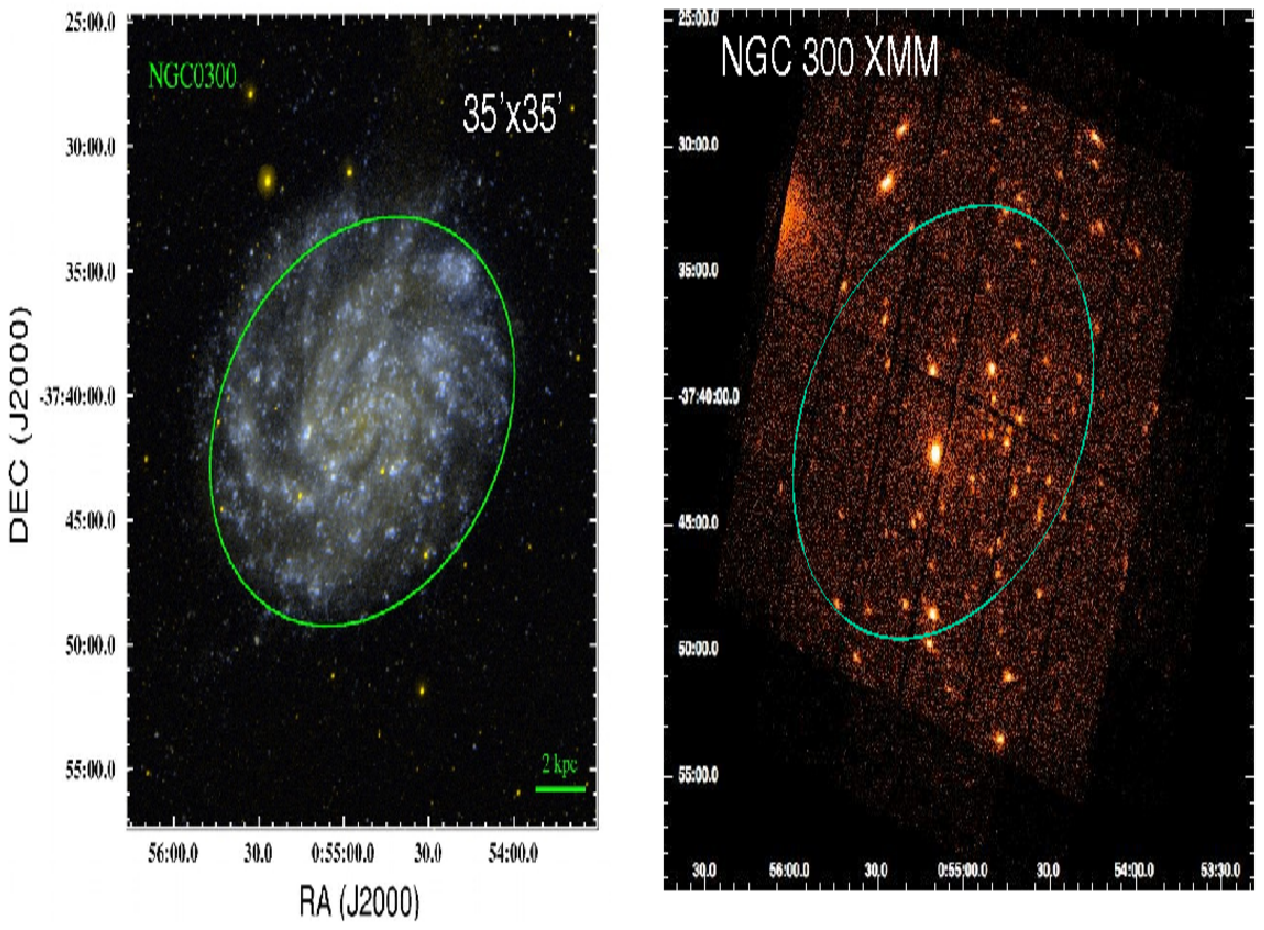}}
\caption{\footnotesize
An example of a face-on galaxy, NGC 300 ({\it left panel}, green ellipse marking D$_{25}$), and its observation with XMM-Newton (EPIC-pn, {\it right panel}).  While the whole galaxy is contained in the FoV, there is little left to explore the area outside D$_{25}$. 
}
\label{ngc300}
\end{figure*}

The field of view provided by WFXT has the fundamental advantage over current instruments that it can cover the large angular size of nearby galaxies with a limited number of separate pointings, providing at the same time enough space for a proper sampling of the immediate neighbouring environment. For example M31, the nearest spiral to us, would need $\sim 10$ pointings, compared to  $\sim$  30 needed with XMM-Newton to cover it out to D25 only.  The observations will include also at least two of its satellites -M32 and NGC 205- and enough area outside the disk to study the outer halo regions and the local environment.
M33, NGC 253,  NGC~300 (see Fig.~\ref{ngc300}), NGC~1291, M106 are all examples of galaxies with sizable angular sizes, that can be covered in a single field together with enough area for a local determination of the background. 

As shown by the example in Fig.~\ref{ngc300}, bright sources can be easily resolved and studied individually.  The depth at which they will be resolved will depend upon the final spatial resolution reached for WFXT, since crowding will be a real issue even in nearby systems, and will be different in different environments. Being able to observe the entire system and its surrounding in each observation 
will have a significant impact e.g. in building long-term light curves of all sources in a given galaxy for variability studies, which in turns has an impact on our understanding of their nature, when spectral or spatial information are not sufficient.

However, to exploit the advantages of WFXT will require careful considerations of the strategy of the  observations, both to study  individual sources  (spectral studies, XLF, time variability, source classification, etc) and to detect and characterize the unresolved emission and the galaxy outskirts and environment.  In particular, in order to be able to exploit the advantage of a large FoV and use a local determination of the background, one major requirement is that we know the instrumental response {\it very well}, as a function of energy and position on the detector. 

One didactic example is given by the extremely elaborate analysis that was required to measure the characteristics of the halo in NGC 253  \cite[see][for details of the analysis required]{Bauer}. The size of the halo and its low surface brightness posed a real challenge to the available XMM-Newton data, both because the source almost fills the FoV, making a local determination of the intensity and spectral shape of the background quite difficult, and because the calibration of the instrument response, required to extrapolate the background values to the area covered by the emission, was not accurate enough.  If we want  to fully exploit the  potential of WFXT to measure low surface brightness extended features, proper calibration issues must be considered.

\section{Large samples}

Current studies of galaxies suffer from small sample statistics, since only a few normal galaxies are typically detected in X-ray surveys \citep{Tajer07, Georgant05, Hornschemeier05, Brusa10}.
The huge increment in the number of sources which will be possible
with WFXT has incredible potential in improving our knowledge in many areas, provided that the
identification of the X-ray sources is done in an automated but secure
way. One of the main issues will be to be able to resolve the small fraction of normal galaxies from the much larger populations of AGNs. Although some discrimination can be made on the basis of the X/O ratios \cite[e.g., as done since ][]{macc}, low redshift galaxies  have the advantage that their X-ray extent can 
be used as a classification tool: at z$ \leq 0.2$ the WFXT PSF should still be able to show them as extended sources, and might allow us to distinguish those dominated by a bright nuclear source.  We also expect them to be optically bright, thus easily recognizable on most  plates.

\subsection{X-ray luminosity functions}
\label{XLF}
As discussed in Ranalli (this conference) the number of objects in the
three WFXT surveys will be so large that we can derive a logN-logS and
a XLF (after identification!!) for different types of galaxies and at different
redshifts. This will allow us to put a more firm base for cosmological studies. Based on our current knowledge of the XLF of normal galaxies, we expect that  the density of galaxies,  at the medium survey flux, is
almost 10/deg$^2$, for a grand total of about 30 000 galaxies at the end
of the survey!

\subsection{ULX}
\label{ULX}
Provided they are covered, we will be able to count on a very large number of photons for sources already known and studied. This will allow a precise determination of their
spectral shape, which in some cases cannot be determined with current data. With the large statistics we can study their
temporal behaviour, to better assess their nature,  and investigate the presence of characteristic
variability patterns (QPOs, PSD etc) which might help in determining
the mass of the compact object.

We also expect to be able to discover many more of these sources, and to extend their detection and characterization to objects at much larger z, to study them in different contexts.  It is well known that ULXs are invariably associated with star formation activity and young population of stars.  However,  we need to quantify what percentage of ULX can be associated with compact objects in binary systems and what other phenomena might give rise to the extreme luminosities observed.  At the present time,  we expect that about
a quarter of ULXs are SNR, but this is based on a very scanty statistics
\citep{Swartz04}. Possibly the different species of ULXs can be discriminated
on the basis of their spectral shape. For instance a thermal model
would be preferentially associated to a SN, while classical binary
are mostly defined by power laws, modified by a disk and/or corona
in a few cases. If covered at the sensitivity of the medium survey, WFXT will provide enough photons to distinguish spectra of many ULXs
up to the distance for example of the Cartwheel, which hosts a large population of ULXs and one of the brightest known to date \citep{wt, wtc}.   The lack of a well determined spectral shape is hampering our understanding of the nature of even the brightest one, which can be interpreted as a compact, $\sim$100 M$_{\odot}$ BH binary or a SN \citep{pizzo}. 

\section{Hot gas in the local universe}
\label{gas}

\begin{figure}[ht!]
\resizebox{\hsize}{!}
{\includegraphics[clip=,width=3.5cm]{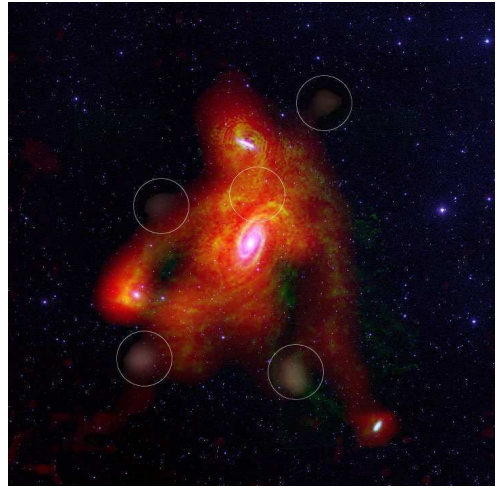} 
}
\caption{\footnotesize
Composite radio optical image of the M81/M82 galaxy group, covering an area of  $\rm \sim 3^o\times 3^o$ \cite[from][]{Chynoweth}
}
\label{HI}
\end{figure}

HI envelops and high velocity cloud systems (HVC) 
are observed around  spiral galaxies  (e.g. M83, NGC 2403)
and between objects
(NGC4631 connecting to its companions, \citealt{rand};
M81 connecting to
NGC3077/M82, \citealt{Yun94}). These HI halos extend over large angular scales and appear to be dominated by filamentary structures (see e.g. Fig~\ref{HI})
clearly demonstrating 
disruption of the system by tidal  interactions. Some HVCs may be the remnants of galaxy formation currently being accreted  or may be  associated to current episodes of vigorous star formation like supernovae explosions. A hot halo is also detected in these same objects, so we need to understand whether both come from the same
source. The presence of extend halos could also indicate the existence of large reservoirs of gas at low surface brightness and large size that could go undetected.

Similarly, a connection is expected between early type galaxies and their surroundings. The evidence of cavities in the diffuse X-ray surface brightness in clusters of galaxies points to a close connection with a central AGN radio outbursts. In X-ray bright clusters, cavities are prominent and can be easily identified. This phenomenon is now beginning to emerge 
at all scales, down to bright central group galaxies and normal galaxies \citep[see][for a recent discussion of this topic]{diehl}.  However, due to the smaller scale and the lower luminosities expected, detections of cavities is increasingly more difficult in groups and galaxies. And yet they are expected to play a more  prominent role in the evolution of the host structure, given the shallower potential and the faster dynamical time-scales than in clusters.

\begin{figure}[ht!]
\resizebox{\hsize}{!}
{\includegraphics[clip=,width=3.5cm]{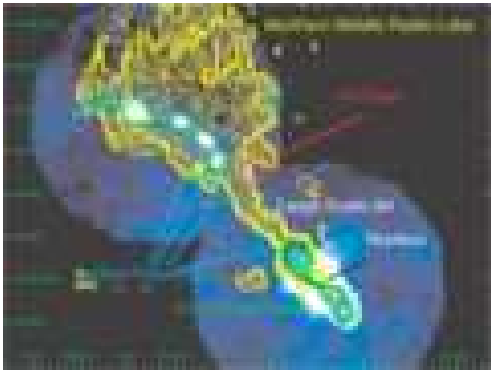} 
}
\caption{\footnotesize
Exposure corrected, Gaussian smoothed XMM-Newton image of Cen A in the 0.3-1.0 keV band pass
\cite[from][]{kraft}. The field displayed is 40$'$ in size and does not include all of the diffuse emission of CenA and its interesting features.
}
\label{cena}
\end{figure}

We can benefit from observations of nearby objects, such as Centaurus A, which exhibits complex morphology on spatial scales from milliarcseconds to degrees (Fig~\ref{cena}), to understand the relation and the interference between jets and environment, feedback etc.   Observing the local universe, where powerful radio sources exist, allows us to obtain details and  relatively large fluxes --- but the angular sizes require the large and effective FoV that WFXT can provide. 

\section{Conclusions}
Although not in the primary science, the exploration of the local universe is possible and in fact almost mandatory to fully exploit the WFXT performance.
We have shown a few cases in which just the large focused area of WFXT would
greatly improve our knowledge. The size of the samples of all kinds of galaxies and of individual sources will be much larger than has been possible so far, opening new areas of exploration and allowing us to confirm tentative results based on small number statistics.
It is important to focus on the observation strategy, 
in order to include nearby objects of crucial interest, on the identification procedure,  to rapidly have a list of candidates, and on proper calibration of the whole FoV, to fully exploit it.
Repeated observations of the same area of sky will yield also a base for
variability studies, which can be vital for the identification and classification of many sources.

\vskip -0.5truecm
\bibliographystyle{aa}

\end{document}